**An improved peridynamic framework to eliminate unphysical stress and fictitious yield at geometry surface for geomaterials' elastoplastic deformation and fracture analysis**

Yixin Li[a], Xueyu Geng[a,*]

[a] School of Engineering, University of Warwick, CV4 7AL, Coventry, UK

* Corresponding author.
E-mail address: Xueyu.Geng@warwick.ac.uk (Professor at the University of Warwick)

**Abstract**: This paper presents an improved non-ordinary state-based peridynamics (NOSB-PD) framework for modelling the elastoplastic behaviour and damage of geomaterials, such as soil, rock, and concrete, under quasi-static conditions. Conventional NOSB-PD for elastoplastic materials faces two primary challenges: the surface effect due to the low accuracy of the approximate deformation gradient ($F^{PD}$) near boundaries and fictitious yielding during explicit time integration. These issues can lead to numerical errors, such as inaccurate crack predictions and potential simulation failure. The proposed framework thoroughly analyses the reason for the surface effect by demonstrating that $F^{PD}$ exhibits only first-order accuracy within a horizon radius ($\delta$) from the surface but introduces residual stresses within a larger range of $2\delta$ with significant surface effect. To mitigate this, a divergence formulation of the non-local differential operator (NDO) is applied to enforce a reasonable stress gradient within a $2\delta$ subregion, alongside a traction boundary condition consistent with the divergence of stress. Additionally, a "loading-balance-correction" algorithm is introduced to enhance the conventional explicit time integration process. The model is validated by integrating it with a modified hyperbolic-hardening Drucker-Prager model, where the stress integration is performed using the closest point projection method (CMMP). Numerical results, compared with finite element simulations and experimental data, demonstrate the effective elimination of the surface effect and false yielding, providing a robust simulation of elastoplastic deformation and progressive failure in geomaterials.

# 1. Introduction

Soil, rock and concrete are pressure-dependent materials that undergo complex elastoplastic deformation [1], and they are commonly used in large-scale geotechnical structures such as foundations, slopes, tunnels, and geothermal facilities. Given the potential economic losses and casualties caused by damage to these critical infrastructures [2, 3], it is essential to understand the finite deformation and progressive failure of these geomaterials to design more resilient infrastructures.

Research has developed various numerical methods to simulate the deformation of geotechnical structures, but many face challenges in accurately tracing crack initiation and propagation along arbitrary paths [4]. In conventional continuum mechanics (CCM), physical quantities like stress and strain are defined through spatial differentials due to the CCM's local nature [5]. However, this continuity assumption breaks down at cracks [6, 7]. In this regard, the finite element method (FEM) has become a robust tool for modelling the mechanical response of geomaterials under specific boundary conditions [8], but it requires additional techniques like remeshing to handle mesh distortion or discontinuities [9, 10]. While the extended finite element method (XFEM) allows for mesh breaking, it struggles with accurately modelling the coalescence and branching of multiple cracks [11, 12]. Alternatively, traditional mesh-free methods like the Particle Flow Code (PFC) and Discrete Element Method (DEM) have been explored to solve deformation and failure problems. However, they face difficulties in tracking irregular cracks based on particle clusters and require complex considerations of particle size effects and the derivation of particle interaction mechanisms from the material's constitutive laws [4,13-15].

To address these limitations from the conventional numerical methods, Silling [16] introduced peridynamics (PD), a mesh-free method based on a non-local kernel. In this framework, the partial differential equations of the CCM are reformulated into a spatial integral and temporal differentiation form [17]. Each material point interacts with its neighbouring points within a horizon through non-local forces [18-20]. Damage is quantified by the fraction of broken bonds, thereby overcoming the singularity at crack tips and enabling simulation of spontaneous crack propagation [21]. The earliest version of PD, known as bond-based PD (BB-PD), models interactions solely between connected points but imposes constraints such as a fixed Poisson's ratio and plastic incompressibility [22]. While these limitations are improved in ordinary state-based PD (OSB-PD), its parameters lack clear physical meaning, making experimental calibration difficult. Additionally, elastoplasticity analysis in OSB-PD requires reformulating constitutive models into a non-local framework, limiting its adaptability to various material models [23, 24]. To facilitate strain analysis and incorporate classical constitutive laws straightforwardly into the framework of the PD, Silling et al. [25] proposed a peridynamic correspondence material model in the non-ordinary state-based PD (NOSB-PD) framework. This approach approximates the deformation gradient, enabling PD internal forces to be computed directly from existing CCM constitutive laws. While NOSB-PD has been successfully applied to linear elastic and hyperplastic materials [4, 9, 26], its application to elastoplastic materials introduces two key challenges that cause numerical errors.

First, incomplete horizons at boundary points lead to unstable spatial integration, resulting in irregular material stiffness and sharp discontinuities in the deformation field, an issue known as the surface effect in PD [27]. Some research overlooked this error unless the unphysical displacement becomes significant [24, 28]. The unphysical forces induced by this surface effect can lead to false cracks, ultimately compromising the simulation [27]. Various approaches have been developed to mitigate the surface effect issue. One intuitive solution is to extend virtual material layers beyond the computational region to create complete horizons for the boundary points [29]. However, this method alters the geometric characteristics. The size and displacement of artificial layers require extra trial-and-error tests, especially for the roll support and the traction load [19, 27, 30]. An alternative strategy introduces a surface correction coefficient based on energy equivalence principles [31]. The mirror nodes method provides a standardised procedure for creating artificial layers, but it poses challenges in determining mirror particles in the complex geometric configurations, including those material points at corners [32, 33]. Beyond artificial layers, nonlocal differential operators, such as the peridynamic differential operator (PDDO) and the general nonlocal differential operator (NDO), help bridge the gap between the local and nonlocal theories

[34]. In this context, Madenci et al. [27] and Yu et al. [35] reformulated the governing equation of PD to directly apply the displacement constraints and external traction to the outmost material layer, effectively eliminating the surface effect. Another approach involves local-nonlocal coupling, which integrates PD boundary conditions with those in conventional continuum mechanics (CCM) [36, 37]. Yu et al. [38, 39] and Chen et al. [40] introduced a novel functional layer at the boundary, which makes it possible to couple the PD boundary conditions with those in CCM, whereas residual unphysical force persists at the interface.

The second major numerical challenge in elastoplastic materials is false yielding, which arises when solving PD's equation of motion using explicit time integration [41]. Near loading points, material particles may undergo premature yielding and plastic deformation before the computational domain reaches global equilibrium. In materials that experience plastic fracture, false yielding can lead to unrealistic crack formation [42, 43]. To mitigate this issue, Zhou et al. [41] initially treated boundary particles as elastic and later transitioned them to elastoplastic. However, the duration of the initial elastic phase cannot be determined in advance. An alternative approach iteratively solves the PD equilibrium equation using an implicit scheme, updating the incremental displacement field [27, 38, 44-46]. While implicit methods construct a global Jacobian matrix to solve large, sparse systems, they demand significant computational memory (RAM). Recent advancements in GPU hardware and high-performance computing (HPC) clusters have enabled efficient parallelisation, making explicit time integration more attractive [47, 23]. In the NOSB-PD framework, once a new stress field is computed, material point displacements can be updated efficiently using parallel computing cores. Therefore, improving explicit time integration is crucial to preventing false yielding in elastoplastic materials.

Considering these challenges, this study develops an improved NOSB-PD with an explicit time integration scheme for analysing rate-independent elastoplastic deformation and failure, in which the surface effect and false yielding are effectively removed. The paper is structured as follows: Section 2 introduces the fundamental NOSB-PD theory. Section 3 analyses boundary effects and derives an improved governing equation to eliminate surface errors. Section 4 presents a "loading-balance-correction" algorithm for the explicit time integration to remove fictitious yielding. Section 5 validates the proposed PD model. Finally, the concluding remarks are presented.

## 2. Formations of the conventional NOSB-PD

### 2.1. Kinematics

In this section, we clarify the fundamental variables used to describe the kinematics in NOSB-PD. As depicted in Fig. 1 (a), the reference (initial) and deformed configurations are denoted by $\Omega_0$ and $\Omega_t$, respectively. They are mapped by a motion, $\psi_t$. The total Lagrange approach is used to describe deformation behaviours [48].

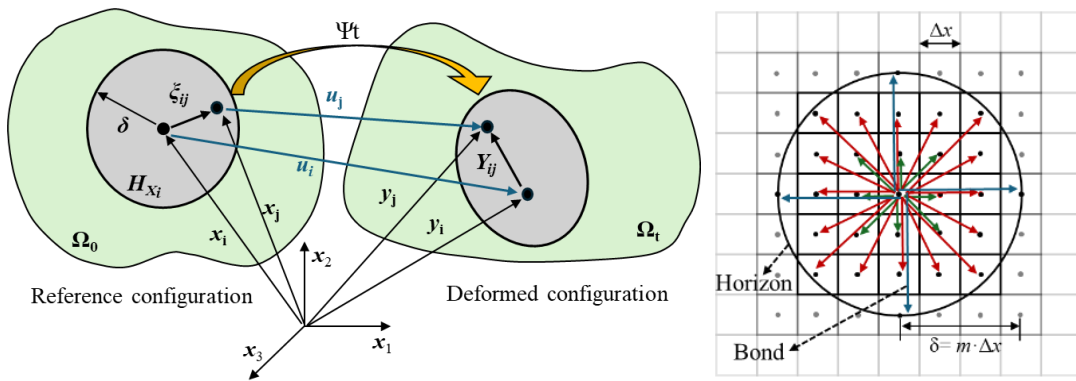

**Fig. 1.** Schematics of (a) non-local interaction within a horizon and (b) the deformation of bonds and the mapping relationship between reference and deformed configurations.

In NOSB-PD, the computational region is discretised as a certain number of material points. As illustrated in Fig. 1 (b),

material point ($x_i$) interacts with a set of family members ($x_j$) located in the horizon $H_i$. The connection between two material points is defined as a bond. In $\Omega_0$, the relative position vector is calculated by $\xi_{ij} = x_j - x_i$. In $\Omega_t$, material points $x_i$ and $x_j$ move to a new position of $y_i$ and $y_j$, respectively. The relative displacement vector is $\eta_{ij} = (y_j - x_j) - (y_i - x_i)$. NOSB-PD defined a concept of deformation vector state ($\underline{Y}[x_i,t]$) as the basic variables to quantify the deformation at a material point [49]. $\underline{Y}[x_i,t]$ contains all the relative positions for the family points of $x_i$ in $\Omega_t$. Correspondingly, $\underline{X}[x_i,t]$ denotes the initial relative position vector state in $\Omega_0$ as follows (also see Fig. 1):

$$\underline{X}\langle \xi_{ij} \rangle := \underline{Y}[x_i,0]\langle \xi_{ij} \rangle = x_j - x_i, \quad x_j \in H_i = \xi_{ij} \|\xi_{ij}\| \leq \delta \tag{1}$$

$$\underline{Y}[x_i,t]\langle \xi_{ij} \rangle := y_j - y_i = \xi_{ij} + \eta_{ij} = \xi_{ij} + u_j - u_i \tag{2}$$

where the angle bracket denotes that the state notation is related to the bond $\xi_{ij}$. In the CCM, the deformation gradient tensor is used to describe the deformation of an infinitesimal segment, i.e., $F^{CCM} = \partial y / \partial x$. Similarly, a non-local approximate deformation gradient, $F^{PD}$, can be defined to map a cluster of bonds into their deformed patterns:

$$\underline{Y}\langle x_i, t \rangle \langle \xi_{ij} \rangle = F^{PD} \cdot \xi_{ij} \tag{3}$$

To minimise the integrated mapping error for all the bonds of $x_i$, the least squares method can be used to derive the explicit expression of $F^{PD}$, i.e.:

$$\frac{\partial}{\partial F^{PD}}\left[ \int_{H_i} \omega(\|\xi_{ij}\|) \cdot (\underline{Y}[x,t]\langle \xi_{ij} \rangle - F^{PD} \cdot \xi_{ij})^2 dV_j \right] = 0 \tag{4}$$

The explicit expression of $F^{PD}$ (written as $F$ in the following text) reads:

$$F = \left[ \int_{H_i} \omega(\|\xi_{ij}\|) \cdot (\underline{Y}\langle \xi_{ij} \rangle \otimes \xi_{ij}) dV_j \right] K_i^{-1} \tag{5}$$

where $\omega$ is a weight function depending on the length of a bond, $\otimes$ represents the dyadic product, d$V$ denotes the volume of a lattice and $\|\bullet\|$ represents the L$^2$-norm to calculate the length of a bond. The tensor $K$, also known as the non-local shape tensor, is given by:

$$K_i = \int_{H_i} \omega(\|\xi_{ij}\|) \cdot (\xi_{ij} \otimes \xi_{ij}) V_j \tag{6}$$

The constitutive correspondence principle allows one to conduct strain analyses in the framework of NOSB-PD. The approximate deformation gradient can be left polar decomposed to a sequence of a rotation tensor ($R$) followed by a homogeneous left stretch tensor ($V$), i.e., $F=V\cdot R$. An integration strategy can be used to compute $V$ and $R$ iteratively [50]. The rate of $F$ is given by:

$$\dot{F} = [\int_{H_i} \omega(\|\xi_{ij}\|) \cdot (\underline{\dot{Y}}\langle \xi_{ij} \rangle \otimes \xi_{ij}) dV_j] K^{-1} \tag{7}$$

where $\underline{\dot{Y}}$ is the rate of deformation vector state. The velocity gradient $L$ reads:

$$L = \dot{F} F^{-1} \tag{8}$$

which is employed to obtain the stretch rate tensor ($D$) and the rotation rate tensor ($W$) via

$$D_{ij} = \frac{1}{2}(L_{ij} + L_{ji}) \qquad W_{ij} = \frac{1}{2}(L_{ij} - L_{ji}) \tag{9}$$

An antisymmetric rotation tensor can be calculated using the permutation tensor:

$$\Omega_{ij} = e_{ikj}\omega_k \tag{10}$$

with,

$$\begin{cases} \boldsymbol{\omega} = \boldsymbol{w} + (\mathbf{1} \cdot \mathrm{tr}(\boldsymbol{V}) - \boldsymbol{V})^{-1} \boldsymbol{z} \\ z_{ij} = e_{ikj}(DV)_{jk} \\ W_{ij} = e_{ikj} w_k \end{cases} \quad (11)$$

where $\boldsymbol{z}$, $\boldsymbol{w}$ and $\boldsymbol{\omega}$ are auxiliary tensors and $\mathbf{1}$ is the identity matrix. The current rotation tensor $\boldsymbol{R}_t$ and left stretch tensor $\boldsymbol{V}_t$ can be updated by

$$\begin{aligned} \boldsymbol{R}_t &= (\mathbf{1} - 0.5 \cdot \Delta t \boldsymbol{\Omega})^{-1}(\mathbf{1} + 0.5 \cdot \Delta t \boldsymbol{\Omega}) \boldsymbol{R}_{t-\Delta t} \\ \boldsymbol{V}_t &= \boldsymbol{V}_{t-\Delta t} + \Delta t \cdot (\boldsymbol{LV} - \boldsymbol{V\Omega}) \end{aligned} \quad (12)$$

The unrotated deformation rate tensor is obtained by:

$$\boldsymbol{d} = \boldsymbol{R}^{\mathrm{T}} \boldsymbol{D} \boldsymbol{R} \quad (13)$$

The unrotated strain tensor increment can be determined by

$$\Delta \boldsymbol{\varepsilon} = \boldsymbol{d} \cdot \Delta t \quad (14)$$

Incremental unrotated strain tensor is used to compute current unrotated Cauchy stress using a selected material constitutive law, which is required to compute PD internal force vector. Details are discussed in Section 4.

2.2. Dynamics

The peridynamic equation of motion at the particle $\boldsymbol{x}_i$ is given by:

$$\rho \ddot{\boldsymbol{u}}(\boldsymbol{x}_i, t) = \boldsymbol{L}(\boldsymbol{x}_i, t) + \boldsymbol{b}(\boldsymbol{x}_i, t) \quad (15)$$

where $\rho$ is the mass density, $\boldsymbol{L}$ is the internal force vector and $\boldsymbol{b}$ is the body force vector. Eq. (15) has been proven to comply with the angular momentum balance and objectivity [51]. In the PD theory, $\boldsymbol{L}$ is no longer computed by the divergence of stress but the integration of bond force density vectors:

$$\boldsymbol{L}(\boldsymbol{x}_i, t) = \int_{H_i} \boldsymbol{f}_i(\boldsymbol{\xi}_{ij}, t) \mathrm{d}V_j = \int_{H_i} (\underline{\boldsymbol{T}}[\boldsymbol{x}_i, t]\langle \boldsymbol{\xi}_{ij} \rangle - \underline{\boldsymbol{T}}[\boldsymbol{x}_j, t]\langle \boldsymbol{\xi}_{ji} \rangle) \mathrm{d}V_j \quad (16)$$

where $\boldsymbol{f}_i$ is the pairwise force density vector and $\underline{\boldsymbol{T}}[\boldsymbol{x}_i, t]$ is the force state, a collection of bond forces connected with particle $\boldsymbol{x}_i$. Two pairs of work-conjugated quantities are used to derive the expression of $\underline{\boldsymbol{T}}[\boldsymbol{x}_i, t]$ [25, 52]:

$$\Delta W^{\mathrm{PD}} = \int_{H_i} (\underline{\boldsymbol{T}}\langle \boldsymbol{\xi}_{ij} \rangle)^{\mathrm{T}} \cdot \Delta \underline{\boldsymbol{Y}}\langle \boldsymbol{\xi}_{ij} \rangle \mathrm{d}V_j \quad (17)$$

$$\Delta W^{\mathrm{CCM}} = \mathrm{tr}(\boldsymbol{P} \cdot \Delta \boldsymbol{F}^{\mathrm{CCM}}) = \mathrm{tr}(\boldsymbol{P} \cdot \Delta \boldsymbol{F}) \quad (18)$$

where $\Delta W$ is the incremental strain energy density, tr (•) is the trace of a square matrix and $\boldsymbol{P}$ is the first Piola-Kirchhoff (PK) stress tensor. By applying $\Delta W^{\mathrm{PD}} = \Delta W^{\mathrm{CCM}}$ and combining Eq. (5), one obtains:

$$\underline{\boldsymbol{T}}[\boldsymbol{x}_i, t]\langle \boldsymbol{\xi}_{ij} \rangle = \omega(\|\boldsymbol{\xi}_{ij}\|) \boldsymbol{P}_i \boldsymbol{K}_i^{-1} \boldsymbol{\xi}_{ij} \quad (19)$$

2.3. Elimination of zero-energy mode

Recalling Eq. (3), a uniform $\boldsymbol{F}$ is applied to all the associated bonds to compute a deformation vector state. Thereby, the computed $\psi_t(\boldsymbol{\xi}_{ij})$ has a weak dependency on a single bond's performance. Given that the first PK stress is obtained based on a particle's deformation, the mapping between the deformation pattern and the obtained force state is non-unique, which can result in oscillation on displacement field, i.e., zero-energy mode or pseudo-energy mode. This numerical instability can be suppressed by adding a stabilisation term to $\boldsymbol{L}^s$ [53]. First, a nonuniform deformation state is defined to quantify the bond-level mapping error:

$$\underline{z}\langle \xi_{ij}\rangle = \underline{Y}\langle \xi_{ij}\rangle - F\cdot \xi_{ij} \tag{20}$$

If the elastic modulus is constant, an additional force state reads [54]:

$$\underline{T}^s[x_i,t]\langle \xi_{ij}\rangle = \frac{3E}{\pi\delta^4(1-2\nu)}\omega(\|\xi_{ij}\|)\frac{\xi\otimes\xi}{\|\xi\|^3}\underline{z}\langle \xi_{ij}\rangle \tag{21}$$

in which $E$ is Young's modulus and $\nu$ is the Poisson's ratio. The supplementary force vector at particle $x_i$ is formulated as:

$$L^s(x_i,t) = \int_{H_i}(\underline{T}^s[x_i,t]\langle \xi_{ij}\rangle - \underline{T}^s[x_j,t]\langle \xi_{ji}\rangle)dV_j \tag{22}$$

Combining Eq. (15) and Eq. (22), the complete equation of motion of conventional NOSB-PD is given by:

$$\rho\ddot{u}(x_i,t) = \underbrace{L(x_i,t) + b(x_i,t)}_{\text{Conventional PD force}} + \underbrace{L^s(x_i,t)}_{\text{Stablisation force}} \tag{23}$$

## 3. Improved NOSB-PD without surface effect

### 3.1. Completion of the horizon

In the deformed configuration $\Omega_t$, the position of an arbitrary family particle $x_j$ of particle $x_i$ can be rewritten using Taylor series expansion (TSE) as:

$$\begin{aligned} y_j &= \psi(x_i + \xi_{ij}) = y_i + \nabla y_i\cdot \xi_{ij} + \frac{1}{2}\nabla(\nabla y_i\cdot \xi_{ij})\cdot \xi_{ij} + R_2 \\ &\approx y_i + F_i\cdot \xi_{ij} + \frac{1}{2}(\nabla F_i\cdot \xi_{ij})\cdot \xi_{ij} \end{aligned} \tag{24}$$

where $\nabla$ is the gradient operator and $R_n$ denotes n-th order infinitesimal. Then, combining Eq. (24) with Eq. (3) and comparing with Eq. (5), it leads to:

$$\begin{aligned} \left[\int_{H_i}\omega_{ij}(y_j - y_i)\otimes \xi_{ij}dV_j\right]K_i^{-1} &= F_i\cdot\left[\left(\int_{H_i}\omega_{ij}\xi_{ij}\otimes \xi_{ij}dV_j\right)K_i^{-1}\right] \\ &+ \frac{1}{2}\nabla F_i\cdot\left(\int_{H_i}\omega_{ij}\xi_{ij}\cdot \xi_{ij}\otimes \xi_{ij}dV_j\right)K_i^{-1} \end{aligned} \tag{25}$$

In the reference configuration, $F$ is an identity matrix and $\underline{Y}\langle\xi_{ij}\rangle = \xi_{ij}$, i.e., $\left(\int_{H_i}\omega_{ij}\xi_{ij}\otimes\xi_{ij}dV_j\right)K_i^{-1} = \mathbf{1}$. Eq. (25) yields:

$$F_i = \left\{\left[\int_{H_i}\omega_{ij}(y_j - y_i)\otimes \xi_{ij}dV_j\right] - \frac{1}{2}\nabla F_i\cdot\left(\int_{H_i}\omega_{ij}\xi_{ij}\cdot \xi_{ij}\otimes \xi_{ij}dV_j\right)\right\}K_i^{-1} + R_2 \tag{26}$$

Therefore, for those materials possessing an intact horizon (within the dash curve in Fig. 2), the second term in the right-hand side of Eq. (26) is a vacant matrix due to $\int_{H_i}\omega_{ij}\xi_{ij}\otimes \xi_{ij}dV_j = \mathbf{0}$. The approximate deformation gradient computed by Eq. (5) achieves second-order accuracy. However, for material points near the edges, Eq. (5) has only first-order accuracy, which leads to non-physical kinks in the displacement field.

As illustrated in Fig. 2, the computational domain is divided into two subregions by a closed offset curve that maintains a constant distance of $2\delta$ to the boundary. For an arbitrary material point $x_i \in \mathbf{D}$, it possesses a shape tensor identical to that of any family point, i.e., $K_i = K_j = K_0$. This is because the particle distribution is centrosymmetric in an intact horizon. In contrast, for those particles $x_i \in \mathbf{E}$, they have $K_i \neq K_j$. We rephrase the first PK stress using first-order Taylor series expansion (TSE), i.e.: $P_j = P_i + \nabla P_i\xi_{ij} + R_1(\xi_{ij})$. Ignoring the higher-order terms, Eq.(16) is further written as:

$$\begin{aligned}
\boldsymbol{L}^{\mathrm{PD}}(\boldsymbol{x}_i,\mathrm{t}) &= \int_{H_i} \omega(\|\boldsymbol{\xi}_{ij}\|)\cdot(2\boldsymbol{P}_i + \nabla\boldsymbol{P}_i\cdot\boldsymbol{\xi}_{ij})\boldsymbol{K}_0^{-1}\boldsymbol{\xi}_{ij}\mathrm{d}V_j \\
&= 2\boldsymbol{P}_i\boldsymbol{K}_0^{-1}\int_{H_i}\omega(\|\boldsymbol{\xi}_{ij}\|)\boldsymbol{\xi}_{ij}\cdot\mathrm{d}V_j + \int_{H_i}\omega(\|\boldsymbol{\xi}_{ij}\|)\nabla\boldsymbol{P}_i\cdot\boldsymbol{\xi}_{ij}\boldsymbol{K}_0^{-1}\boldsymbol{\xi}_{ij}\mathrm{d}V_j \\
&= \boldsymbol{0} + \int_{H_i}\omega(\|\boldsymbol{\xi}_{ij}\|)\nabla\boldsymbol{P}_i\cdot\boldsymbol{\xi}_{ij}\boldsymbol{K}_0^{-1}\boldsymbol{\xi}_{ij}\mathrm{d}V_j \\
&= \nabla\boldsymbol{P}_i\;,\quad \boldsymbol{x}_i\in D
\end{aligned} \tag{27}$$

and

$$\begin{aligned}
\boldsymbol{L}^{\mathrm{PD}}(\boldsymbol{x}_i,\mathrm{t}) &= \int_{H_i}\omega(\|\boldsymbol{\xi}_{ij}\|)\cdot(\boldsymbol{P}_i\boldsymbol{K}_i^{-1} + \boldsymbol{P}_j\boldsymbol{K}_j^{-1})\boldsymbol{\xi}_{ij}\mathrm{d}V_j \\
&= \int_{H_i}\omega(\|\boldsymbol{\xi}_{ij}\|)\cdot(\boldsymbol{P}_i\boldsymbol{K}_i^{-1} + \boldsymbol{P}_i\boldsymbol{K}_j^{-1} + \nabla\boldsymbol{P}_i\boldsymbol{\xi}_{ij}\boldsymbol{K}_j^{-1})\boldsymbol{\xi}_{ij}\mathrm{d}V_j \\
&\neq \nabla\boldsymbol{P}_i\;,\;\boldsymbol{x}_i\in E
\end{aligned} \tag{28}$$

Recalling the internal force term in the CCM: $\boldsymbol{L}^{\mathrm{CCM}}(\boldsymbol{x}_i,\mathrm{t}) = \nabla\cdot\boldsymbol{P}(\boldsymbol{x}_i,\mathrm{t})$, it is obvious that Eq. (27) satisfies the condition of a smooth stress field. But Eq. (28) indicates unphysical residual stress exists in the subregion $E$, the range of which is larger than that of the unstable displacement field shown in Eq.(26).

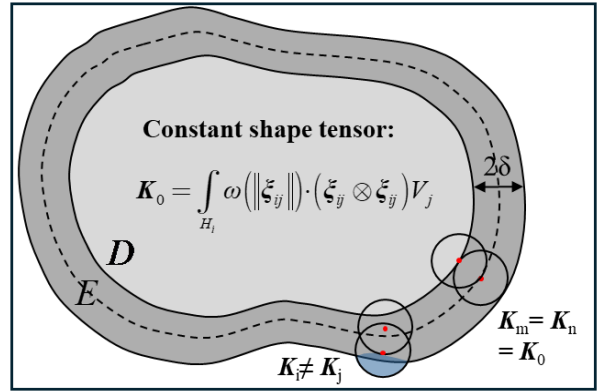

Fig. 2. Schematic of subregions of computational domain.

*3.2. Improved NOSB-PD model*

Comparing Eq.(27) with Eq. (28), a reasonable divergence of stress is required to mitigate the surface effect at the material points in the subregion $E$. On this premise, we used the NDO to enforce a divergence of stress without any unphysical stress [55]:

$$\nabla\cdot(\bullet)_i = \int_{H_i}\omega(\|\boldsymbol{\xi}_{ij}\|)\cdot((\bullet)_j - (\bullet)_i)\cdot(\boldsymbol{K}_i^{-1}\boldsymbol{\xi}_{ij})\mathrm{d}V_j \tag{29}$$

where $(\bullet)$ denotes a physical quantity in the computational domain. The PD internal force vector in subregion $\boldsymbol{E}$ is reformulated as:

$$\begin{aligned}
\boldsymbol{L}^{\mathrm{PD}} &= \int_{H_i}\omega(\|\boldsymbol{\xi}_{ij}\|)\cdot(\boldsymbol{P}_j - \boldsymbol{P}_i)\cdot(\boldsymbol{K}_i^{-1}\boldsymbol{\xi}_{ij})\mathrm{d}V_j \\
&= \int_{H_i}\omega(\|\boldsymbol{\xi}_{ij}\|)\cdot(\boldsymbol{P}_i + \nabla\boldsymbol{P}_i\cdot\boldsymbol{\xi}_{ij} - \boldsymbol{P}_i)\cdot(\boldsymbol{K}_i^{-1}\boldsymbol{\xi}_{ij})\mathrm{d}V_j \\
&= \int_{H_i}\omega(\|\boldsymbol{\xi}_{ij}\|)\nabla\boldsymbol{P}_i\cdot\boldsymbol{\xi}_{ij}\cdot(\boldsymbol{K}_i^{-1}\boldsymbol{\xi}_{ij})\mathrm{d}V_j \\
&= \nabla\boldsymbol{P}_i\;,\;\boldsymbol{x}_i\in E
\end{aligned} \tag{30}$$

which is consistent with the scenario in the CCM, i.e., the surface effect can be effectively removed to ensure a smooth stress field near the boundary.

**Remark 1.** Given Eq. (19) is derived using Eq. (5), the surface effect caused by an inaccurate $F$ is actually in a region that is $2\delta$ to the boundary, although it has been proven that $F$ has second order accuracy within an interlayer of $\delta$ - $2\delta$. In this improved NOSB-PD model, the deformation at boundary particles remains first-order accurate. Residual stress is effectively removed by enforcing a correct stress gradient using NDO. Therefore, the displacement constraints, can be directly imposed on the outmost particle layer.

*3.3. Traction boundary condition*

The external traction (gradient boundary condition) is not naturally included in the PD equation of motion due to PD's nonlocal nature. The conventional NOSB-PD applies external tractions through an equivalent body force vector. In the improved NOSB-PD, the PD internal force vector $L$ is reformulated by NDO, which is much stricter on the Neumann boundary. In the CCM, the balance equation at the traction boundary is $F_{ext} = T_{ext} - P \cdot n = 0$, in which $n$ is the outward normal vector of the boundary. To introduce the net external force to the PD system, the gradient of $F_{ext}$ for the outmost material layer is required to be added to the right hand side in Eq.(29), given by:

$$\nabla \cdot P_i = \int_{H_i} \omega(\|\xi_{ij}\|) \cdot (P_j - P_i) \cdot (K_i^{-1} \xi_{ij}) dV_j + \frac{T_{ext} - P \cdot n}{\Delta x} \tag{31}$$

where $\Delta x$ is the thickness of a material layer. Eq.(31) also maintains dimensional consistency. An improved NOSB-PD model can be present as follows:

$$\rho \ddot{u}(x_i, t) = \begin{cases} \underbrace{\int_{H_i} \omega(\|\xi_{ij}\|) \cdot (P_i K_i^{-1} + P_j K_j^{-1}) \cdot \xi_{ij} dV_j + b + L^s}, & x_i \in D \\ \underbrace{\int_{H_i} \omega(\|\xi_{ij}\|) \cdot (P_j - P_i) \cdot (K_i^{-1} \xi_{ij}) dV_j + b + L^s}_{\text{Stabilised PD internal force}} + \underbrace{\frac{T_{ext} - P \cdot n}{\Delta x}}_{\text{Traction}}, & x_i \in E \end{cases} \tag{32}$$

Note that $n$ is nonvanishing only when a material point $x_i$ is located at the surface positions. As shown in Fig. 3, for those material points at the convex corners, $n$ is the summation of the contributions from two surfaces, e.g., $n = (-1,1)^T$ for the particle at the left top corner. And $n$ is vanishing for the concave corner particles (marked blue). These rules can be extended to a three-dimensional scenario.

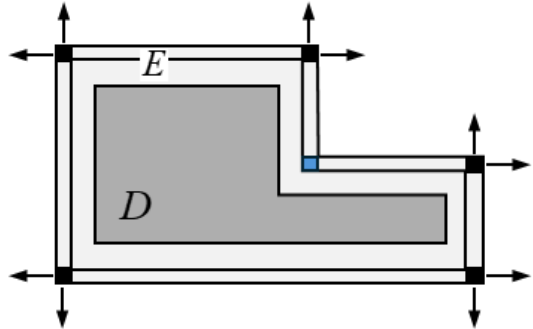

**Fig. 3.** Outward normal vectors at corners.

## 4. Numerical implementations

*4.1. Improved time integration strategy without false yielding*

In the framework of conventional NOSB-PD, boundary material points may prematurely enter the yield state and undergo irreversible plastic deformation before the entire PD system reaches global equilibrium, i.e., false yielding. To address this issue, we developed a "loading-balance-correction" (LBC) algorithm for rate-independent elastoplastic materials. As illustrated in Fig. 4, the computation is conducted in two time-domains. One is the real-time domain, where loadings are applied stepwise (loading module). The other is a virtual time domain (balance module), where the displacement field is iterated in an artificial time domain until the convergence criterion is satisfied:

$$\frac{\|\mathbf{U}_n - \mathbf{U}_{n-1}\|}{\|\mathbf{U}_{n-1}\|} \leq \mathrm{u}_c \tag{33}$$

where $\mathrm{u}_c$ is the tolerance error. At the start of a new loading step, all particles are assumed to behave as linear elastic materials to obtain a trial equivalent state. The computation then proceeds to the plastic correction module, where yielding material points are identified and their constitutive models are updated to be elastoplastic, while other particles retain their elastic properties. If any material point exceeds the yield surface, i.e., $f(\sigma_t, m_t) > 0$, the computation returns to the balance module until a new global equilibrium is achieved, followed by another plastic correction step. As shown in Fig. 4, for the given boundary conditions, the balance and correction modules alternate until all material points remain within their yield surfaces. Subsequently, the old strain tensor and PIV are updated, and the process proceeds to a new loading step.

**Remark 2.** In the "loading-balance-correction" algorithm, global stress, strain increments, and plastic internal variables update using their stabilised values. The balance module ensures a global equilibrium state, while the correction module identifies yielding particles. The loading module applies to a new loading step once global equilibrium is achieved and all particles remain within their yield surfaces. This loading strategy guarantees high accuracy and effectively prevents false yielding. The algorithm can be implemented on the Visual Studio platform with OpenMP for parallel computing to improve computational efficiency.

*4.2. Adaptive dynamic relaxation (ADR) method*

This work employs an adaptive dynamic relaxation (ADR) method to obtain a stable solution with fast convergence under static or quasi-static conditions. A fictitious inertia term and an adaptive local damping coefficient are added in the original equation of motion to dissipate kinetic energy:

$$\mathbf{\Lambda}\ddot{\mathbf{U}}(\mathbf{X},t) + c\mathbf{\Lambda}\dot{\mathbf{U}}(\mathbf{X},t) = \mathbf{F}(\mathbf{U},\mathbf{U}',\mathbf{X},\mathbf{X}') \tag{34}$$

where $\mathbf{\Lambda}$ is the fictitious diagonal density matrix, $c$ is an adaptive damping coefficient determined by the Rayleigh's quotient of the computational system, $\mathbf{X}$ and $\mathbf{U}$ represent the initial coordinate and displacement fields for all the material points and vector $\mathbf{F}$ denotes the summations of forces applied to the PD system. With the central difference method, the global velocity and displacement fields can be explicitly integrated using:

$$\dot{\mathbf{U}}^{n+0.5} = \frac{(2 - c^n \cdot \Delta t)\dot{\mathbf{U}}^{n-0.5} + 2 \cdot \Delta t \mathbf{\Lambda}^{-1}\mathbf{F}^n}{2 + c^n \cdot \Delta t} \tag{35}$$

$$\mathbf{U}^{n+1} = \mathbf{U}^n + \Delta t \cdot \dot{\mathbf{U}}^{n+0.5} \tag{36}$$

For the first iteration step, we can assume the initial conditions for the unknown velocity field at $t^{0.5}$ as:

$$\dot{\mathbf{U}}^{0.5} = \frac{\Delta t \mathbf{\Lambda}^{-1} \mathbf{F}^0}{2} \tag{37}$$

in which $n$ represents the number of iterations and the values of $c^n$ and $\mathbf{\Lambda}$ can be referred to [56].

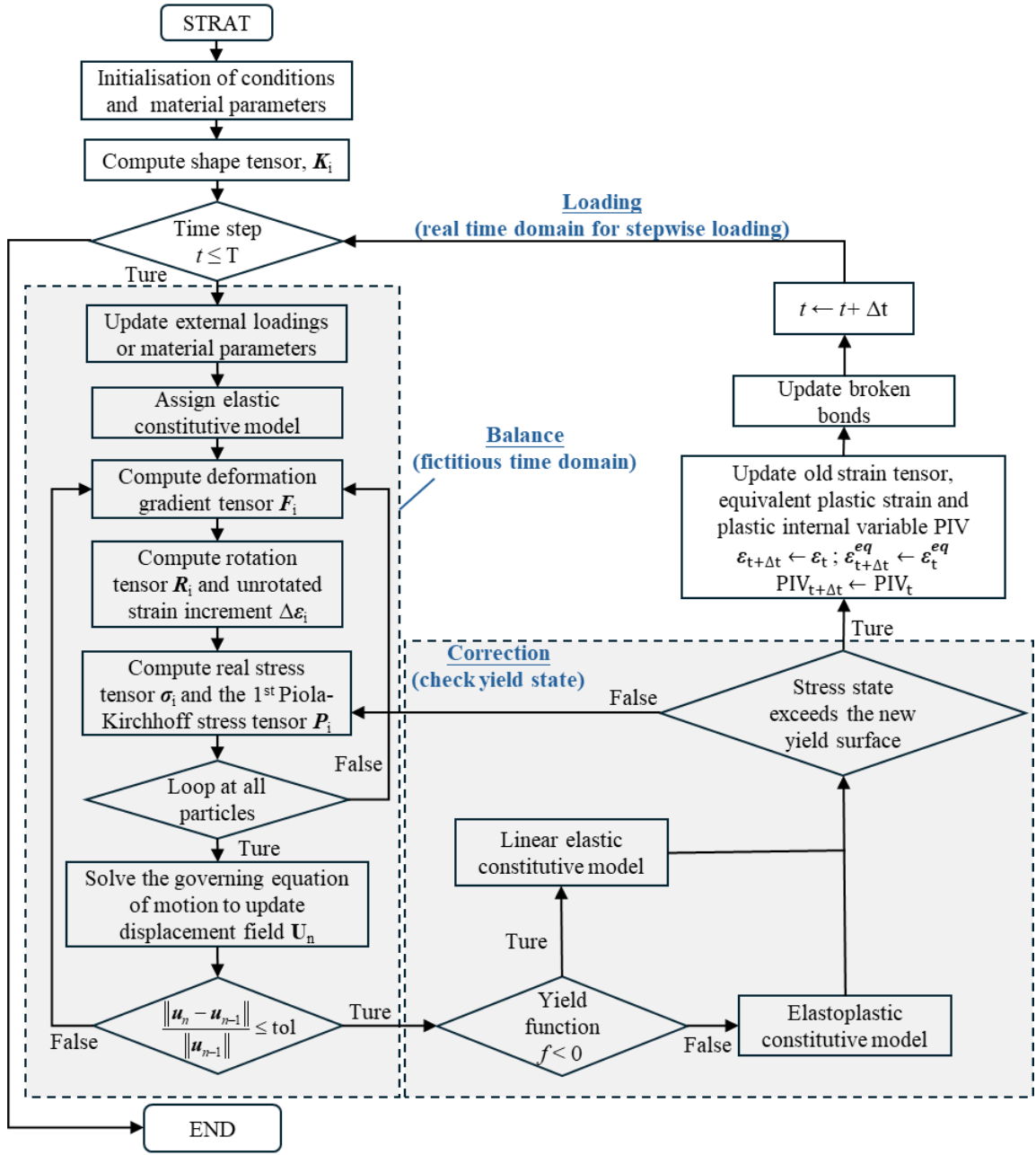

**Fig. 4.** Computational flow chat for improved NOSB PD with elastoplasticity.

*4.3. Drucker-Prager constitutive model*

Drucker-Prager yield criterion is a classical elastoplastic model for pressure-dependent materials such as soil, rock and concrete. It has a smooth yield surface in the octahedral plane (see Fig. 5(a)). The present work employs an extended Drucker-Prager model with two parameters to control the material's plastic flow and the hardening behaviours, respectively [1]. The yield function *f* and plastic potential function *g* are formulated as:

$$\begin{cases} f\left(\boldsymbol{\sigma}, m\left(\xi_p^{eq}\right)\right) = J_2 - mA^\varphi I_1 - mB^\varphi = 0 \\ g\left(\boldsymbol{\sigma}, m\left(\xi_p^{eq}\right)\right) = J_2 - md_0 A^\varphi I_1 - mB^\varphi = 0 \\ A^\varphi = \dfrac{2\sin\varphi}{\sqrt{3}(3-\sin\varphi)}, B^\varphi = \dfrac{6c\cos\varphi}{\sqrt{3}(3-\sin\varphi)} \end{cases} \quad (38)$$

where $I_1$ is the first invariant of the stress tensor, $J_2$ is the second deviatoric stress invariant, $A^\varphi$ and $B^\varphi$ are material parameters determined by internal friction angle $\varphi$ and cohesion $c$, $d_0$ is the dilation parameter ($d_0= 1$ leads to associated flow), $\xi_p^{eq}$ is the equivalent plastic strain working as the plastic internal variable (PIV) and $m$ is the isotropic hardening parameter which can define a hyperbolic hardening law with PIV:

$$m\left(\xi_p^{eq}\right) = m_0 + \frac{h(1-m_0)}{(1-m_0) + h \cdot \xi_p^{eq}} \quad (39)$$

in which $h$ is the hardening parameter. Note that the sign convention in this section is consistent with that in geomechanics, where compressive stress is positive.

To update the stress tensor, one needs to determine if the stress state would exceed the current yield surface for a given strain increment. Assuming the strain increment is purely elastic, trial stress and the yield function are calculated by:

$$\boldsymbol{\sigma}_{n+1}^{tr} = \boldsymbol{\sigma}_n + \boldsymbol{D}^e \Delta\boldsymbol{\varepsilon} \quad (40)$$

$$f_{n+1}^{tr} = f\left(\boldsymbol{\sigma}_{n+1}^{tr}, m_n\right) \quad (41)$$

where $\boldsymbol{D}^e$ is the elastic matrix. If $f_{n+1}^{tr} < 0$, the material remains in the elastic phase. The stress state and the PIV can be directly update by $\boldsymbol{\sigma}_{n+1} = \boldsymbol{\sigma}_{n+1}^{tr}$ and $\xi_{p,n+1}^{eq} = \xi_{p,n}^{eq}$. When $f_{n+1}^{tr} > 0$, the material entre the yield stage, and its deformation becomes elastoplastic. The incremental strain tensor can be divided into the elastic and the plastic parts:

$$\Delta\boldsymbol{\varepsilon} = \Delta\boldsymbol{\varepsilon}^e + \Delta\boldsymbol{\varepsilon}^p \quad (42)$$

in which the plastic strain increment can be calculated based on the flow law:

$$\Delta\boldsymbol{\varepsilon}^p = \Delta\lambda \cdot \frac{\partial g}{\partial \boldsymbol{\sigma}} = \Delta\lambda \cdot \boldsymbol{r} \quad (43)$$

where $\Delta\lambda$ is the incremental consistency parameter. The stress tensor can be updated using:

$$\boldsymbol{\sigma}_{n+1} = \boldsymbol{\sigma}_n + \boldsymbol{D}^e\left(\Delta\boldsymbol{\varepsilon} - \Delta\boldsymbol{\varepsilon}^p\right) = \boldsymbol{\sigma}_{n+1}^{tr} - \Delta\lambda \cdot \boldsymbol{D}^e \boldsymbol{r} \quad (44)$$

According to the consistency condition, the yield surface changes with the stress state; in the meantime, the stress locus is required to be returned to the new yield surface, i.e.,

$$f_{n+1} = f\left(\boldsymbol{\sigma}_{n+1}, m_n\right) = 0 \quad (45)$$

In an isotropic hardening process, the mediate parameter $m$ quantifies the size of the yield surface. From equation (34), the hardening can be expressed as follows:

$$\Delta m = \frac{h \cdot (1-m_0)^2}{\left[(1-m_0) + h \cdot \xi_p^{eq}\right]^2} \cdot \Delta\xi_p^{eq} \quad (46)$$

where the incremental equivalent plastic strain can be calculated by:

$$\Delta \xi_p^{eq} = \sqrt{\frac{1}{2} \|\Delta \varepsilon^d\|} = \Delta \lambda \cdot \sqrt{\frac{1}{2} \|r^d\|} \qquad (47)$$

where $\Delta \varepsilon^d$ and $r^d$ represent the deviatoric part of $\Delta \varepsilon^p$ and $r$, respectively. Combining Eq. (38) to (47), one can solve for the new stress tensor using an implicit return mapping method. In this study, a backward Euler method, closest point projection method (CPPM) is used to update the stress tensor and PIV using an S-space formulation [1]. Details can be found in the appendix.

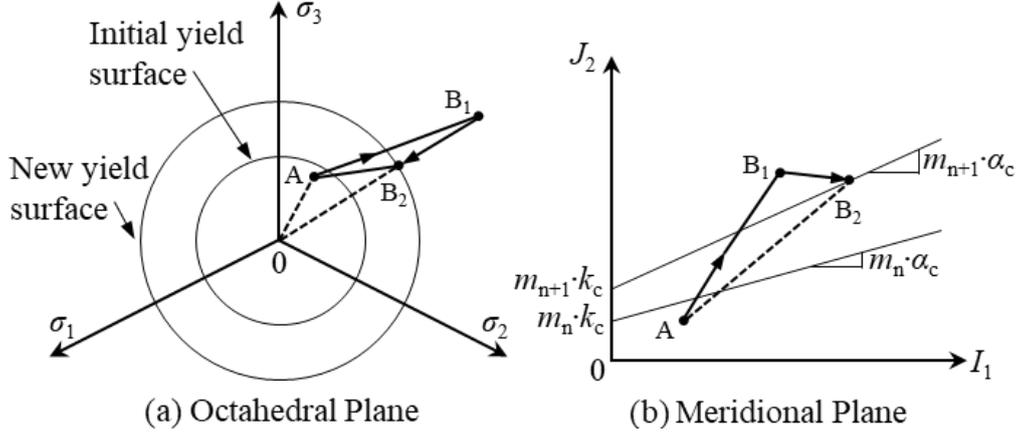

**Fig. 5.** Schematic of Drucker-Prager yield surface.

The updated unrotated stress needs to be rotated back to the true stress tensor (Cauchy stress). The first PK stress tensor **P** in the kinematic equation can be calculated by:

$$\boldsymbol{P} = J\left(\boldsymbol{R}\boldsymbol{\sigma}\boldsymbol{R}^T\right)\boldsymbol{F}^{-T} \qquad (48)$$

where scalar $J$ is the determinant of the deformation gradient tensor.

## 5. Numerical simulation

This section presents four benchmark numerical examples. The first example involves a uniaxial compressive test to validate the convergence characteristics of the proposed model and its ability to eliminate surface effects. The second example simulates a concrete four-point bending test to validate the model's capacity to capture crack propagation. The remaining two simulations focus on the Drucker-Prager constitutive law. In the third example, a series of biaxial compression tests under varying confining pressures were simulated, incorporating different hardening parameters and flow rules, with the results compared to those obtained from FEM simulations. Finally, the plastic strain distribution in a plate with a central hole under compression is analysed to validate plastic strain development using the improved NOSB-PD, particularly near the internal surface.

### 5.1. Validation: Convergence analysis and surface effect removal

As shown in Fig. 6, a rectangular plate with a width of 36 mm and a height of 72 mm in a plane strain case is under a uniaxial compression test. The bottom is fixed in both directions. A uniform stress of 200 kPa is applied to the upper surface. In the PD simulation, the displacement field is initially vacant. The external loading is applied to the outermost material points, linearly increasing until the prescribed magnitude. First, the convergence characteristic of the PD model was checked by assuming the material model is linearly elastic [24, 57]. As shown in Table 1, eight simulations were conducted for $m$-convergence and the $\delta$-convergence analysis. Additionally, the results from FEM, improved NOSB-PD and classical PD were compared to indicate removal of surface effect. The basic material properties are as follows: Young's modulus $E$= 30.0 MPa and Poisson's ratio $v$= 0.25.

Fig. 7 and Fig. 8 display the distribution of displacement and stress along the marked path in the sample (red dashed line). The comparative results from FEM are also plotted. The von Mises stress is a scalar derived from the stress tensor. It can be calculated using the second invariant of the deviatoric stress ($\sqrt{3J_2}$) to reflect the stress state. As shown in Fig. 7 (a) and Fig. 8 (a), when the horizon radius is fixed, the displacement and von Mises stress remain consistent with the FEM results as more material points are included in the horizon. Similarly, Fig. 7 (b) and Fig. 8 (b) demonstrate that when the number of family points remains constant while the horizon size changes, the convergent results are in good agreement with the FEM solution. This indicates that the proposed model satisfies the requirements for $m$-convergence and $\delta$-convergence.

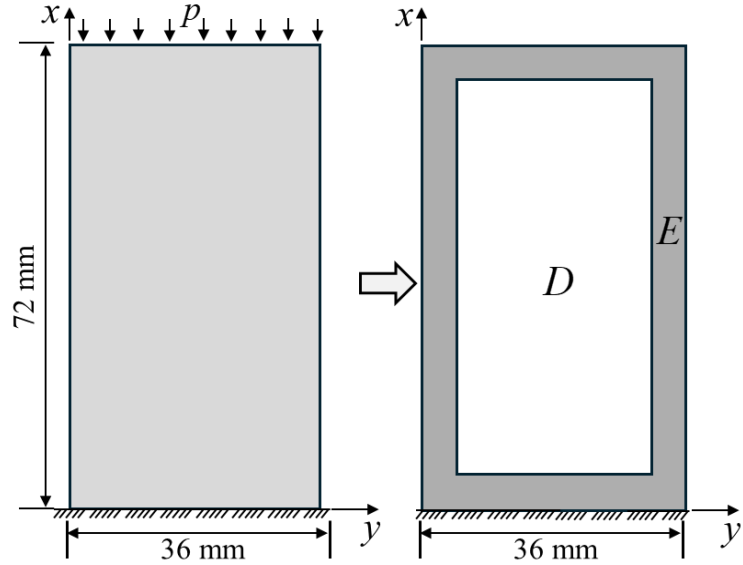

**Fig. 6.** Schematics of (a) Layout of the compression test and (b) improved NOSB PD model.

**Table 1.** Simulation details of convergence study.

|  | $m$-convergence | | | | $\delta$-convergence | | | |
|---|---|---|---|---|---|---|---|---|
| Nonlocal ratio, $m$ | 2.015 | 3.015 | 4.015 | 5.015 | 3.015 | | | |
| Horizon radio, $\delta$ (mm) | 12.0 | | | | 9.0 | 12.0 | 15.0 | 18.0 |
| Particle space, $\Delta x$ (mm) | 6.0 | 4.0 | 3.0 | 2.4 | 3.0 | 4.0 | 5.0 | 6.0 |
| Material points number | 7200 | 16200 | 28800 | 45000 | 28800 | 16200 | 10368 | 7200 |

Fig. 9 plots the von Mises stress contours computed by FEM, classical NOSB-PD, and improved NOSB PD. The models have the same number of elements or material points. The two types of PD models employ identical horizon size. The computational results of improved NOSB PD are consistent with those of FEM, whereas classical NOSB-PD exhibits significantly higher stress values. In addition, a surface effect is observed in classical NOSB-PD results, i.e., significant fluctuations in the stress field near the top surface of the specimen. This error is effectively eliminated by improved NOSB-PD. Furthermore, we investigated the influence of varying horizon sizes on the surface effect. As shown in Fig. 10, the oscillation in the stress field becomes mode significant with an increased horizon size for classical NOSB-PD. The most pronounced fluctuation occurs within around six layers of material points from the traction and displacement-controlled surfaces. This range aligns with the thickness of subregion $E$ in Session 3.

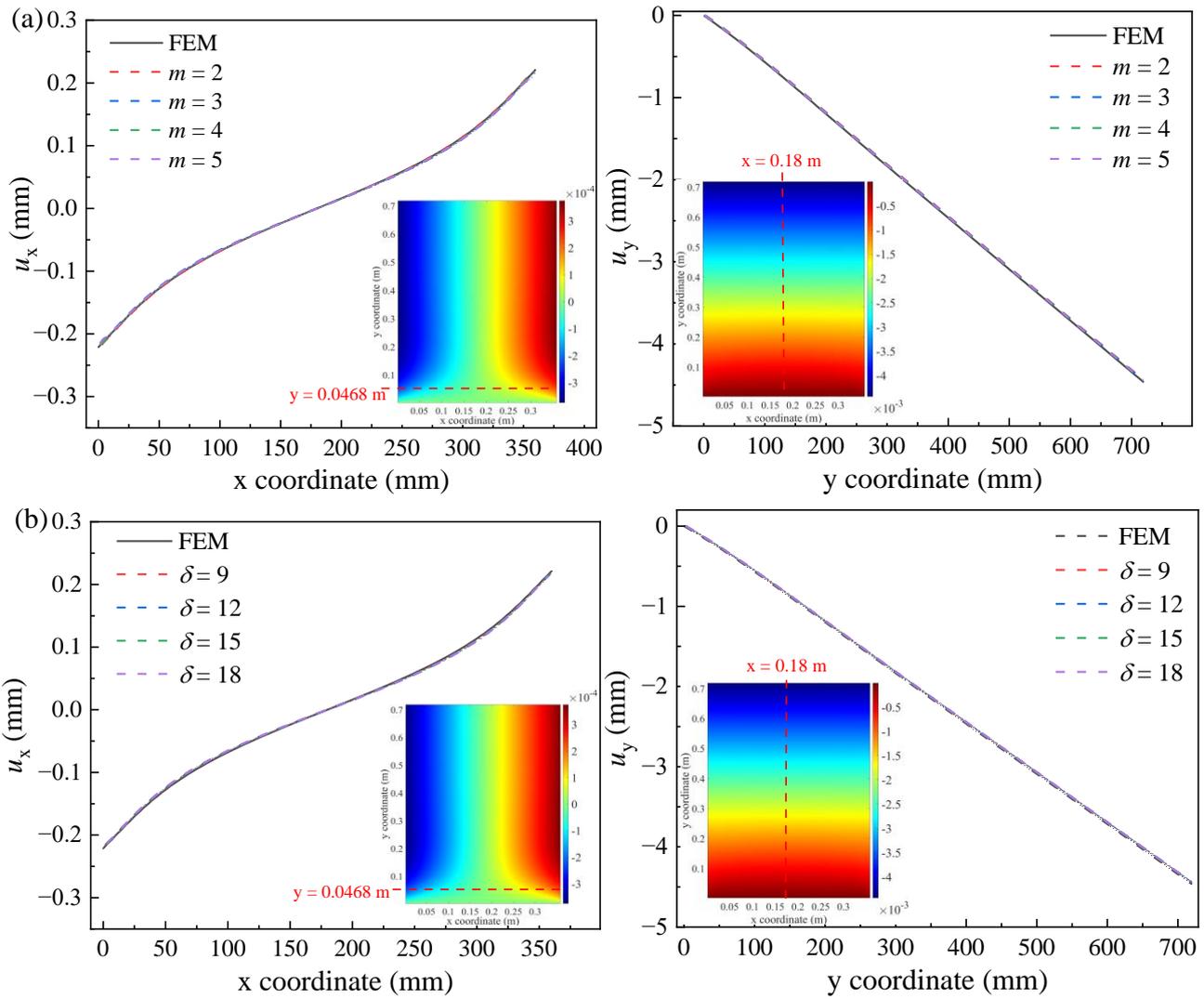

**Fig. 7.** Numerical results of the displacement field: (a) the *m*-convergence analysis and (b) the *δ*-convergence analysis.

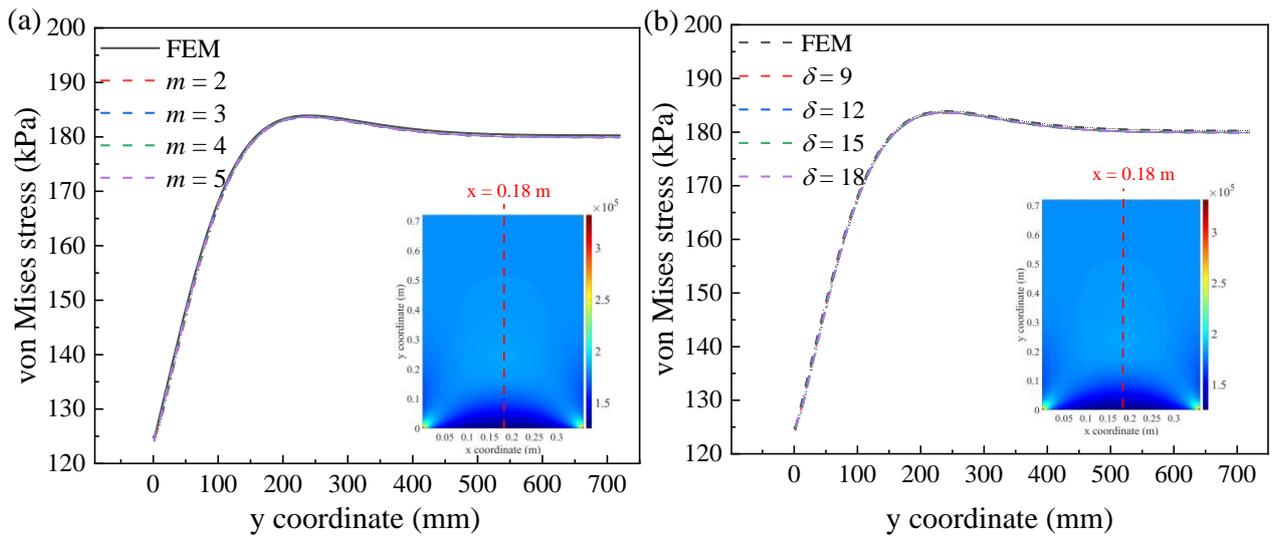

**Fig. 8.** Numerical results of the Mises stress field: (a) the *m*-convergence analysis and (b) the *δ*-convergence analysis.

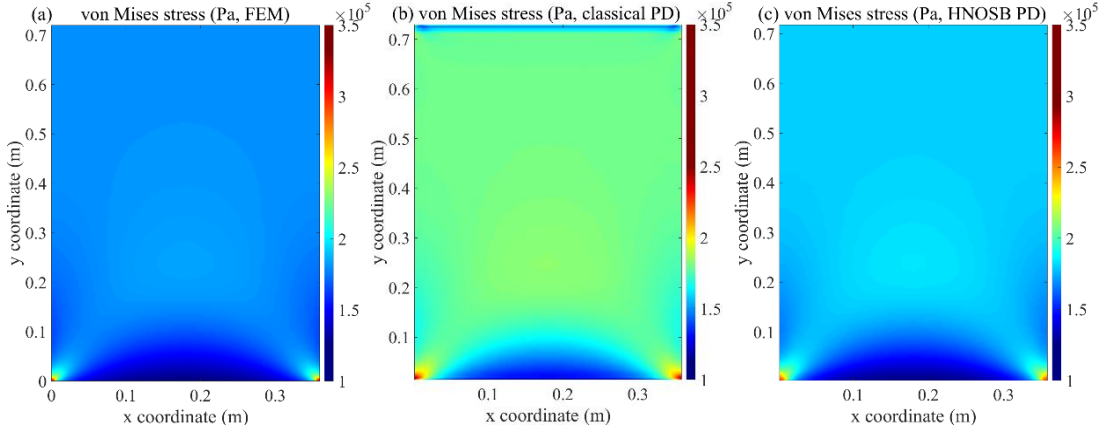

**Fig. 9.** Contours of von Mises stress computed by (a) FEM, (b) classical NOSB PD and (c) improved NOSB PD.

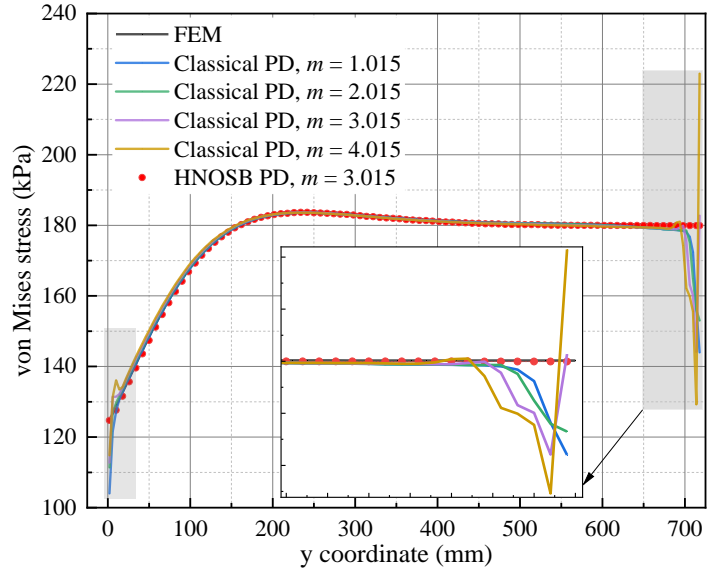

**Fig. 10.** Comparison with the classical non-ordinary state-based peridynamics.

*5.2. Crack propagation of four-point bending test*

In this numerical example, the four-point bending test of a concrete specimen containing a pre-existing flaw is simulated, as well as its loading-displacement curve and crack propagation. The simulation results are compared with experimental data from the literature [58]. The dimensions of the concrete specimen are illustrated in Fig. 11(a). In the PD model, the material points are divided into three layers, including the domain near the pre-existing fissure (see Fig. 11(b)). To simulate the plane stress state of the concrete specimen, 13 layers of material points are arranged in the thickness direction. The specimen is discretised in 25,424 material points with a uniform $\Delta x$ of 2.5 mm. The loading process is carried for 1,000 iteration steps. The Young's modulus is 27 GPa, and the Poisson's ratio is 0.18. In this model, tensile failure is considered. A modified weight function is used to describe a bond's breakage, given by:

$$\omega'(\|\xi_{ij}\|) = \begin{cases} 0, & t_{ij}^{bond} > f_t \\ \omega(\|\xi_{ij}\|), & t_{ij}^{bond} \leq f_t \end{cases} \qquad (49)$$

where $t_{ij}^{bond}$ is the average maximum principal stress at the bond $\xi_{ij}$, $f_t$ is a critical tension stress which is 1MPa in the present model. The variable $t_{ij}^{bond}$ is calculated by:

$$t_{ij}^{bond} = \frac{\sigma_{1,i} + \sigma_{1,j}}{2} \tag{50}$$

where $\sigma_1$ is the maximum principal stress. The cracks are quantified by the fracture of broken bonds at each material point:

$$\vartheta_i = \frac{N_{i,broken}}{N_{i,total}} \tag{51}$$

where $N_{i,broken}$ and $N_{i,total}$ are the number of broken bonds and the total family bond at particle $x_i$, respectively.

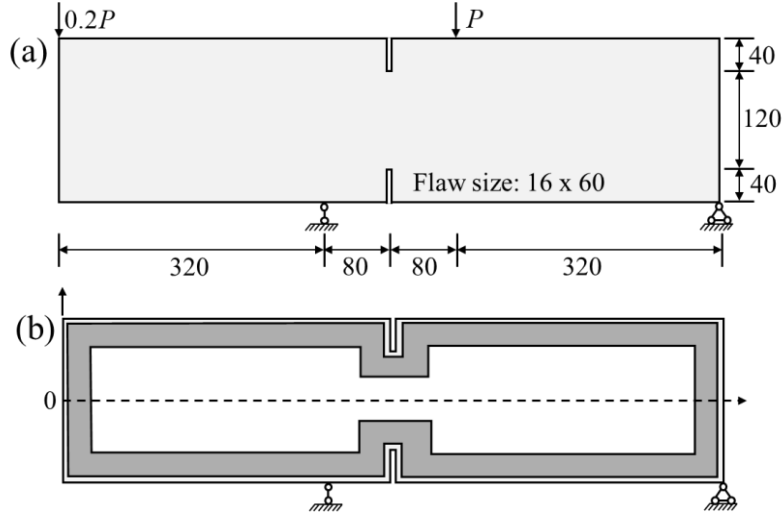

**Fig. 11.** Geometry of the (a) laboratory four-point bending test of the concrete beam with flaws and (b) improved NOSB-PD model (Unit: mm).

The load-displacement curve is presented in Fig. 12. During the initial loading period, the minor fluctuations observed in the PD model may result from insufficient precision in the convergence criterion (*tol*). In the later stages of loading, the maximum load-bearing capacity derived from the PD model shows good agreement with the experimental data. Fig. 13 and Fig. 14 depict the development of cracks and maximum principal stress, respectively. As the load is applied, cracks initially emerge at the right side of the lower fissure and gradually propagate upwards. When the tensile stress on the left side of the upper fissure reaches a critical value, cracks begin to propagate downwards here. Finally, the specimen fails when the two cracks fully penetrate it. As shown in Fig. 14, the maximum tensile stress aligns with the crack propagation path, demonstrating that the PD model effectively captures the influence of bond fracture. Therefore, some conventional failure criteria can be directly incorporated into this PD framework.

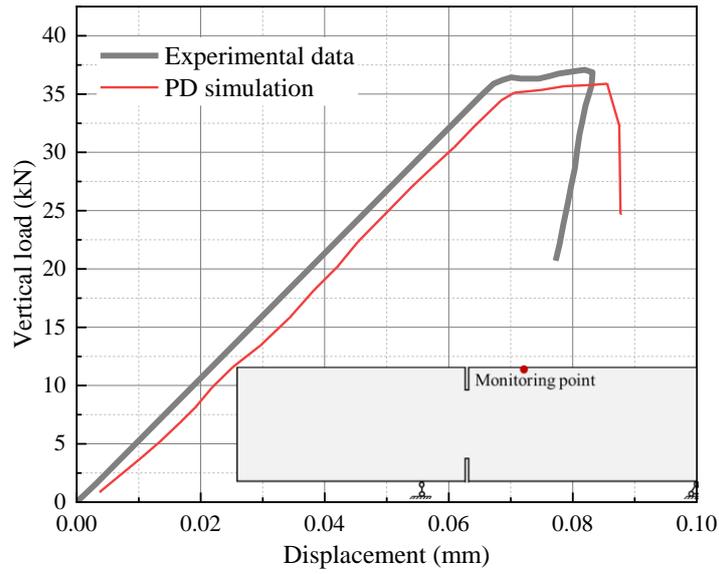

**Fig. 12.** Load-displacement curve in the four-point bending test from literature.

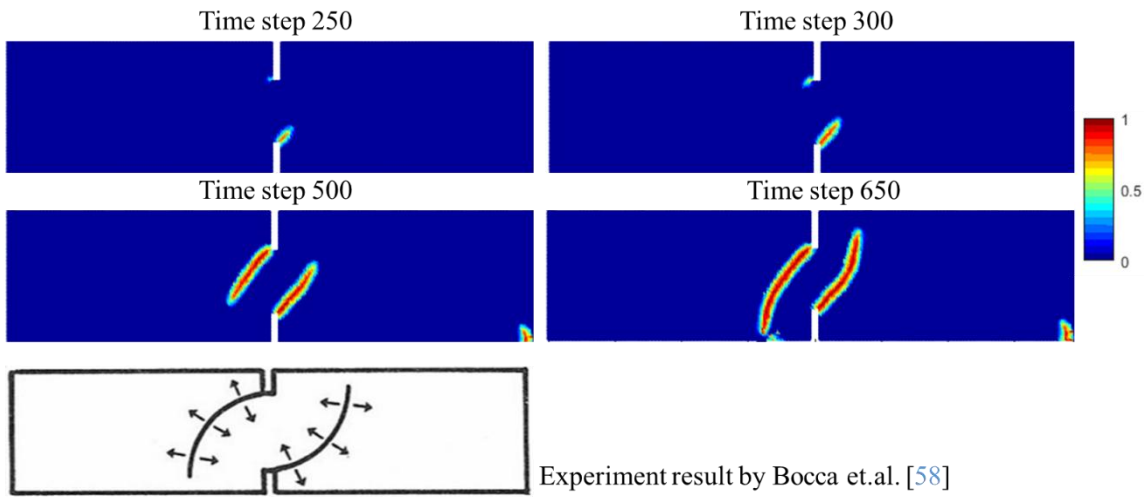

**Fig. 13.** Damage of concrete specimen with two fissures under four-point bending test.

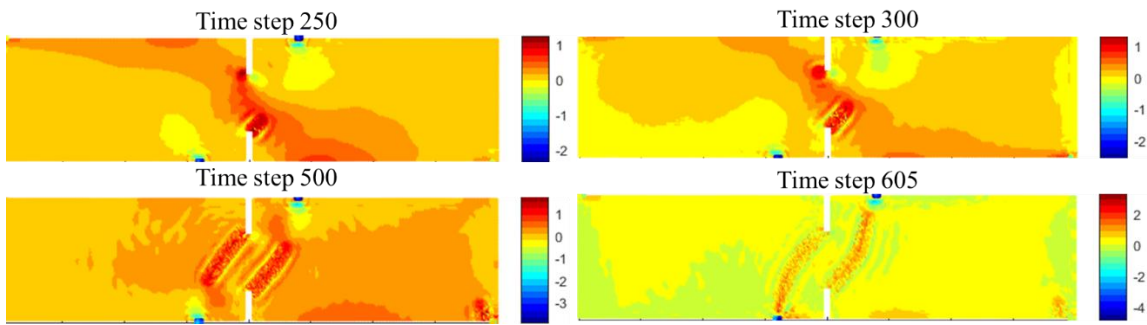

**Fig. 14.** Contour of the maximum principal stress (MPa).

*5.3. Validation: Elastoplasticity without fictitious yield*

5.3.1. Biaxial compression test with different confining stress

To validate the performance of the proposed LBC method in elastoplastic analysis and address the issue of fictitious yield, a sandy loam specimen under plane strain conditions is simulated, as shown in Fig. 15. The specimen is subjected to confining stress at the top and side boundaries. A uniform vertical displacement is applied to the top surface until an axial strain 0.05 is reached, while the bottom is fixed in the vertical direction. The material parameters are: Young's modulus

$E = 30$ MPa, Poisson ratio $v = 0.25$, internal friction angle $\varphi = 35°$ and cohesion $c = 60$ kPa. The loading conditions and constitutive parameters are detailed in Table 2. In Cases 1 to 4, a non-hardening law and non-associated flow rule is adopted ($d_0 = 0$). Cases 6 and 7 investigate the influence of varying hardening parameters. In Case 8, an associated flow rule is employed. In the PD simulation, the traction and displacement boundary conditions were applied linearly. And each of them is configured with 100 iteration steps.

Fig. 16 illustrates the contours of displacement, stress, and equivalent plastic strain for Test 2 at the end of loading. As depicted in Fig. 16(a), under the influence of uniform horizontal force and vertical compression, the displacement field exhibits a smooth distribution without any abrupt changes near the boundaries. In Fig. 16(b), the von Mises stress remains constant (232.1 kPa) throughout the computational domain, further confirming the effective elimination of surface effects. Fig. 16(c) presents the distribution of equivalent plastic strain, which is uniform globally, indicating the absence of false yielding.

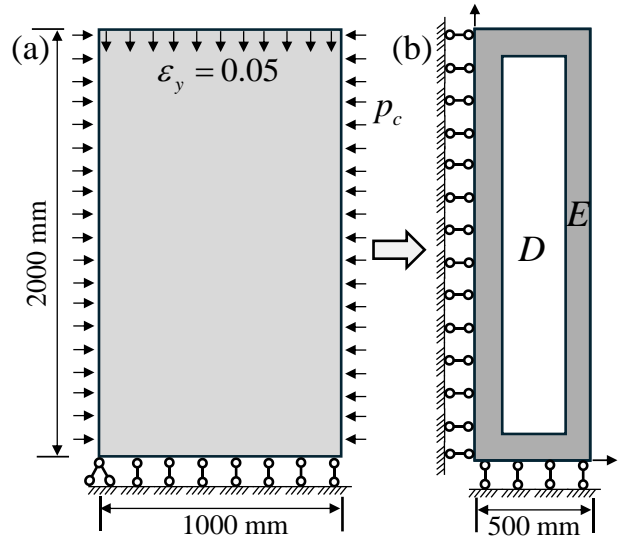

**Fig. 15.** (a) Schematic and boundary of the specimen under biaxial compression and (b) symmetry-simplified improved NOSB-PD model.

**Table 2.** Material parameters of simulation.

| No. | Friction angle (°) | Cohesion(kPa) | $m_0$ | $d_0$ | $h$ | Confining stress (kPa) | Vertical strain |
|---|---|---|---|---|---|---|---|
| 1 |  |  |  |  |  | 400.0 |  |
| 2 |  |  |  |  | 0 | 600.0 |  |
| 3 |  |  |  |  |  | 800.0 |  |
| 4 | 35 | 60.0 | 0.2 | 0 |  | 1000.0 | 0.05 |
| 5 |  |  |  |  | 25 |  |  |
| 6 |  |  |  |  | 50 | 600.0 |  |
| 7 |  |  |  | 1.0 | 0 |  |  |

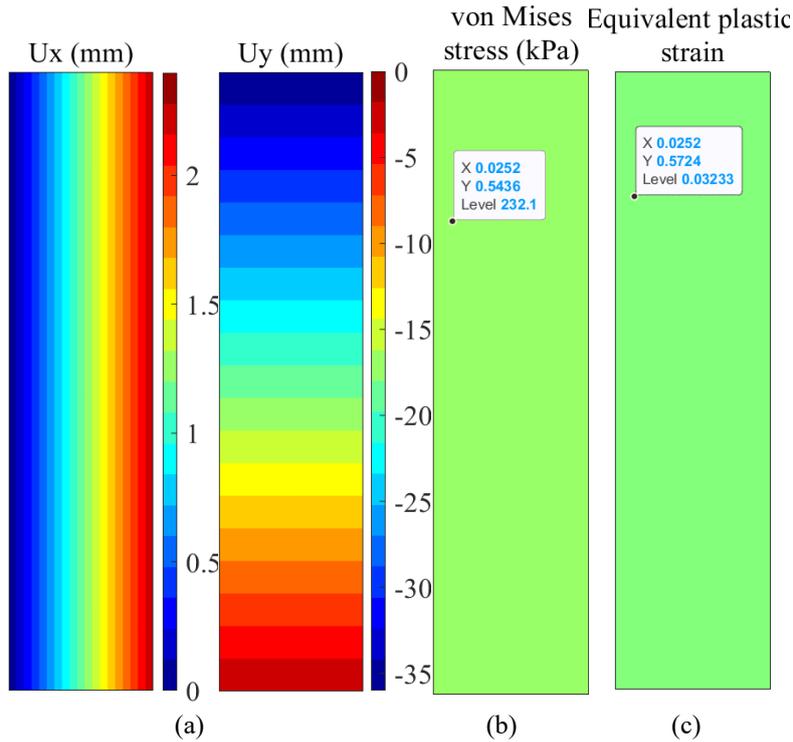

**Fig. 16.** Distribution of displacement, stress and strain fields at the end of simulation (Test 2): (a) displacement, (b) von Mises stress, (c) equivalent plastic strain.

The relationship between von Mises stress and equivalent strain at the centre point of the sample is shown in Fig. 17. As illustrated in Fig. 17(a), during the consolidation phase, the material point initially undergoes an elastic stage, followed by yielding when the equivalent strain reaches 0.0011. After yielding, von Mises stress continuously increases with the rising confining stress. The rate of von Mises stress increase is more pronounced for larger values of the hardening parameter. In the compression phase (Fig. 17(b)), the development of von Mises stress for non-hardening materials (Tests 1-4 and Test 7) gradually stabilises, ultimately exhibiting perfect plasticity, while Tests 5 and 6 demonstrate strain hardening behaviours. Test 7, which considered an associated flow rule, shows a stress-strain relationship identical to that of Test 2, consistent with findings in [1]. The evolution of the equivalent plastic strain is depicted in Fig. 18. With higher confining stress, the sample enters the yield state earlier. For Tests 2, 5, 6, and 7, it is observed that a larger hardening parameter ($h$) results in smaller plastic deformation. Additionally, the associated flow rule (Test 7), which accounts for the dilatancy of plastic deformation, leads to greater plastic deformation. Note that, the computational results of the PD model show good agreement with those of FEM, indicating that the proposed PD model can effectively describe the elastoplastic behaviours of geomaterials and addresses false yielding.

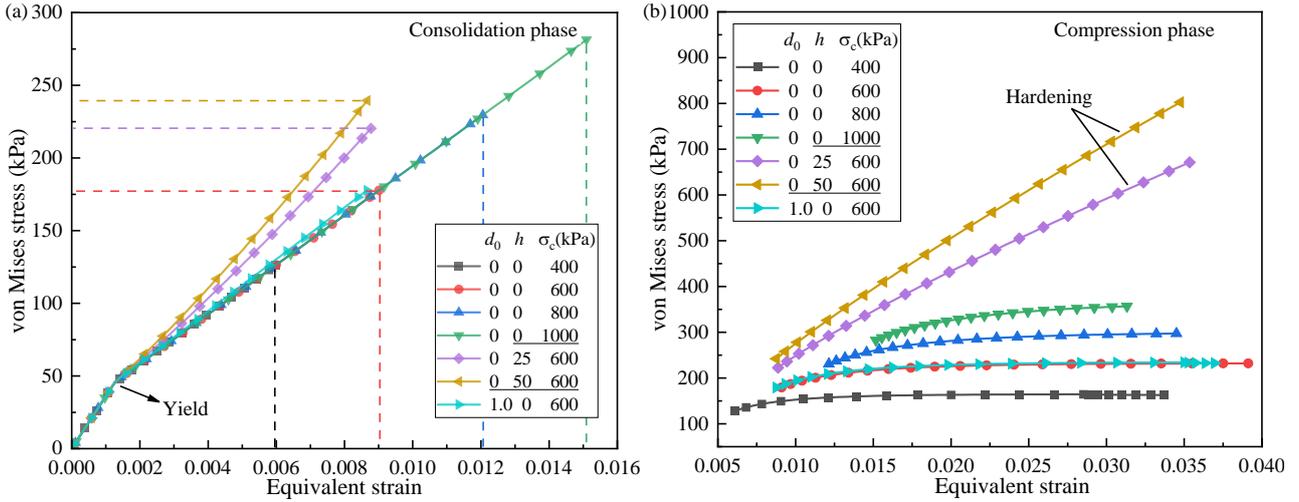

**Fig. 17.** Von Mises stress versus equivalent strain in the (a) consolidation and (b) compression phases (Solid line: FEM results; scatter: PD results).

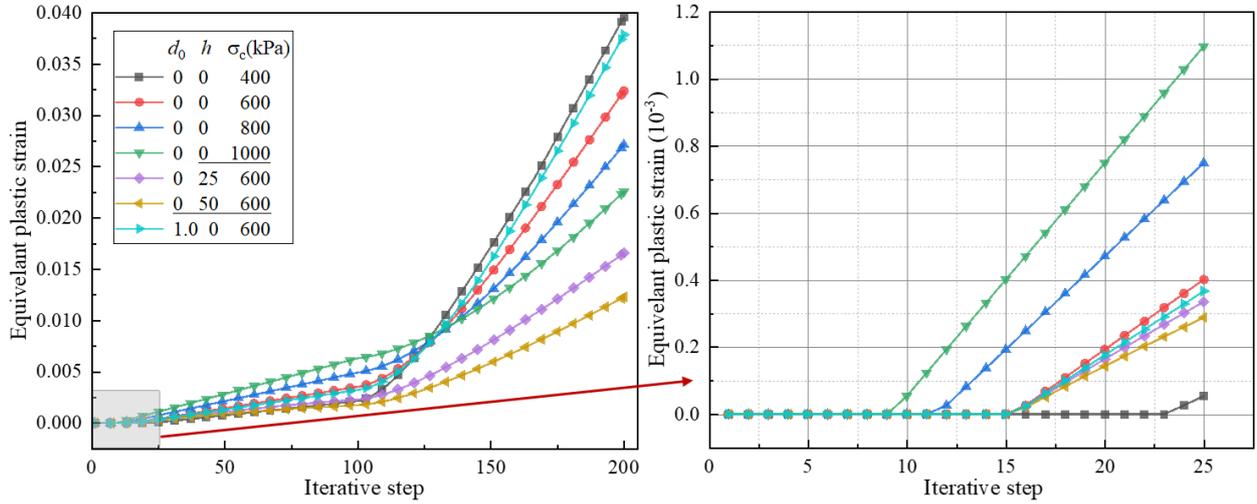

**Fig. 18.** Development of equivalent plastic strain with the external loading (Solid line: FEM results; scatter: PD results).

5.3.2. Compression of a plate with a hollow

In this subsection, the deformation of a plate with a central hollow under uniaxial compression is simulated. The plate has a side length of 100 mm, with a circular hole of 10 mm diameter located at the centre. As shown in Fig. 19 (b), the plate is divided into five subregions in which two boundary layers are set along the four edges and around the central hole, respectively. The material parameters are as follows: Young's modulus of 30 MPa, Poisson's ratio of 0.25, internal friction angle of 35° and cohesion of 60 kPa. A non-associated flow rule is adopted, and the strain hardening is neglected ($h = 0$, $d_0 = 0$). Initially, the stress and strain of the plate are vacant. A uniform displacement of $u_y = 5$ mm is applied to the top and bottom boundaries to induce compressive loading. Both vertical boundaries are free of constraints. The plate is in a plane strain condition.

The computed displacement field is shown in Fig. 20. The displacements near boundaries are smooth without any oscillation occurring. To investigate plastic deformation, three material points were selected: two points are at the left and the upper edges of the hole, respectively, and the third is located on the plate's diagonal. Fig. 21 (a) depicts the evolution of equivalent plastic strain at monitoring points under vertical loading. In the initial loading phase, all three material points maintain elastic behaviour, exhibiting zero equivalent plastic strain. The onset of plasticity occurs at point 1 when the vertical displacement reaches 0.011 mm, followed by point 3 at 0.021 mm. As illustrated in Fig. 21 (b), the von Mises stress is constant with increasing equivalent strain after yielding, suggesting perfect plastic behaviour without strain hardening. Throughout the entire loading process, point 2 remains in the elastic state.

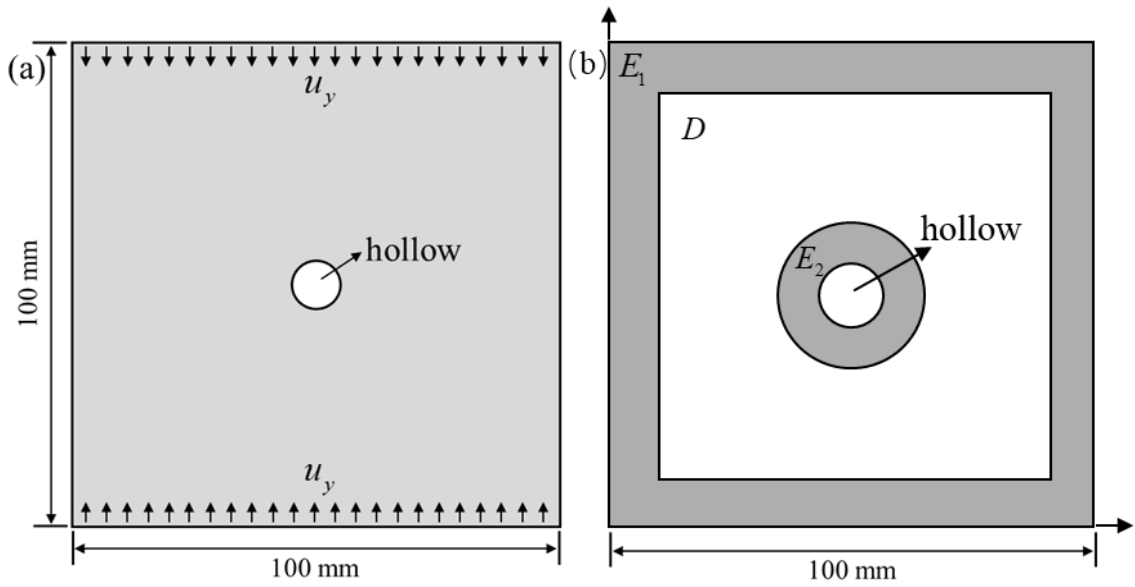

**Fig. 19.** (a) Schematic and boundary of the plate with a hole under compressive load and (b) improved NOSB PD model.

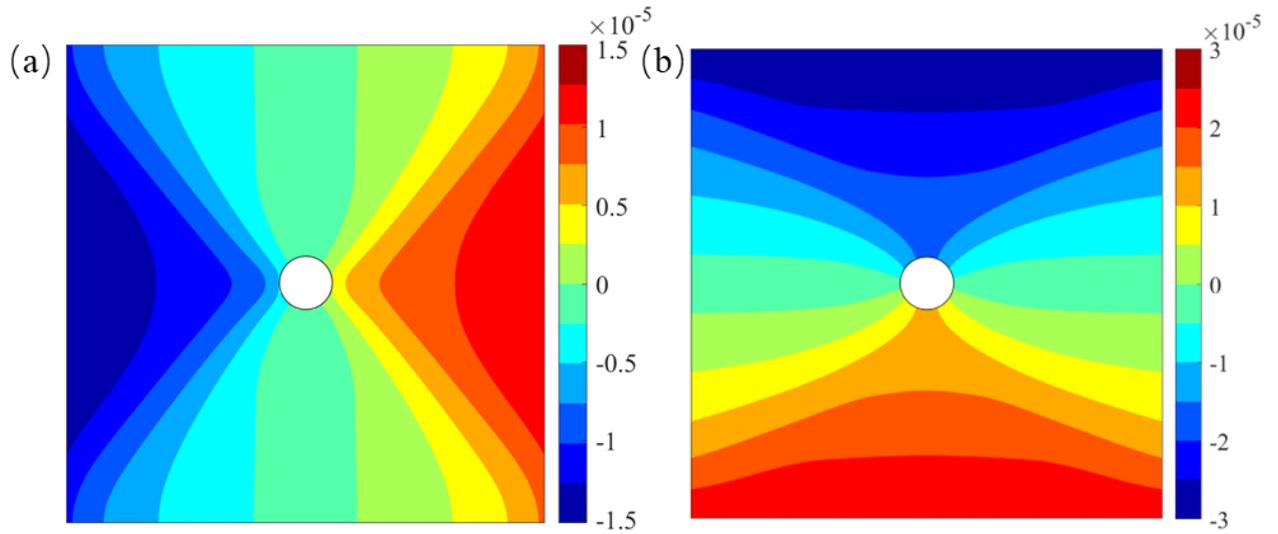

**Fig. 20.** Displacement field computed improved NOSB PD model: (a) horizontal and (b) vertical displacement.

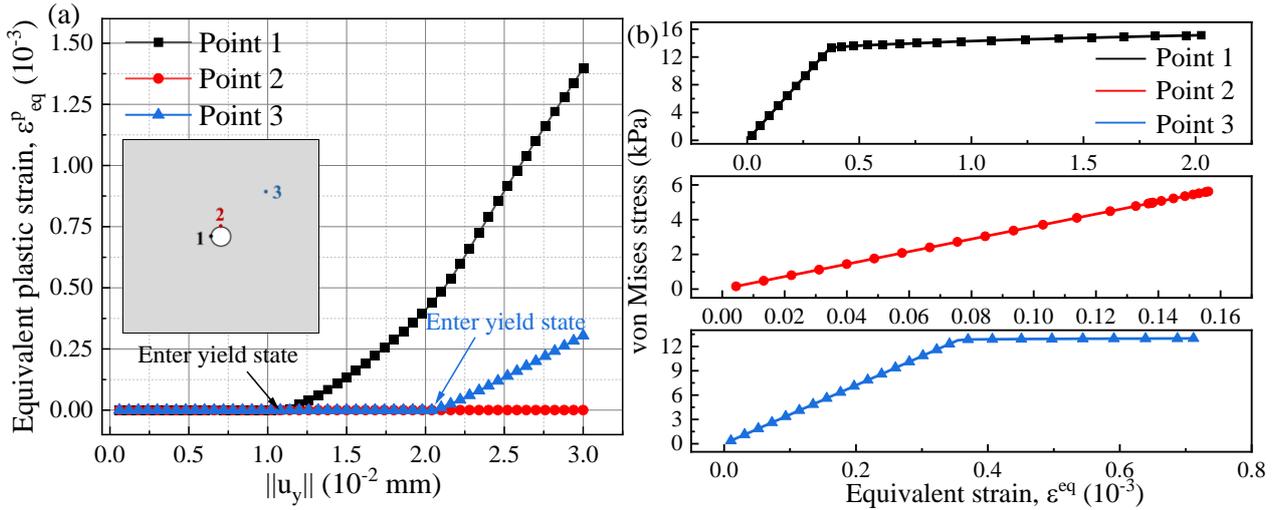

**Fig. 21.** (a) Development of equivalent plastic strain and (b) equivalent stress versus equivalent strain at the marked material points.

Fig. 22 depicts the evolution of equivalent plastic strain when the vertical displacement is 0.0162 mm, 0.018 mm, 0.021 mm and 0.024 mm, respectively. The plastic deformation initiates at the left and right peripheries of the hole and subsequently propagates radially toward the four corners of the plate. The plastic zone progressively evolves into a symmetrical X-shaped shear band, showing clear strain localisation, while the surrounding material remains elastic. As demonstrated in Fig. 23, the spatial distribution of von Mises stress shows a good correlation with the progression of plastic regions. This confirms that the material undergoes shear-dominated yielding behaviour without numerical errors by fictitious yield.

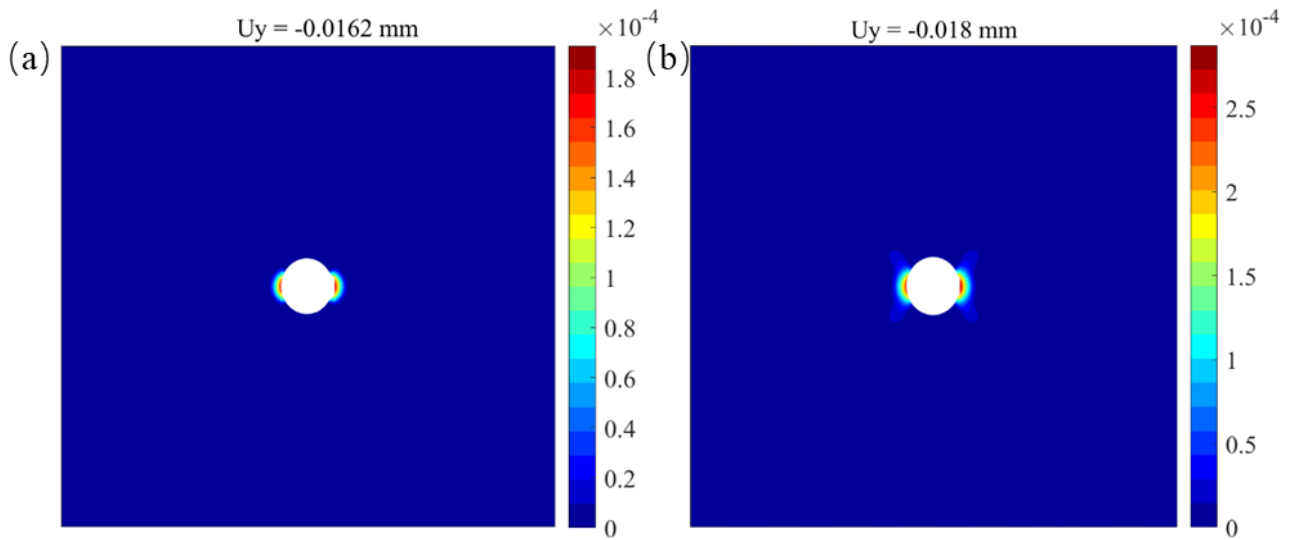

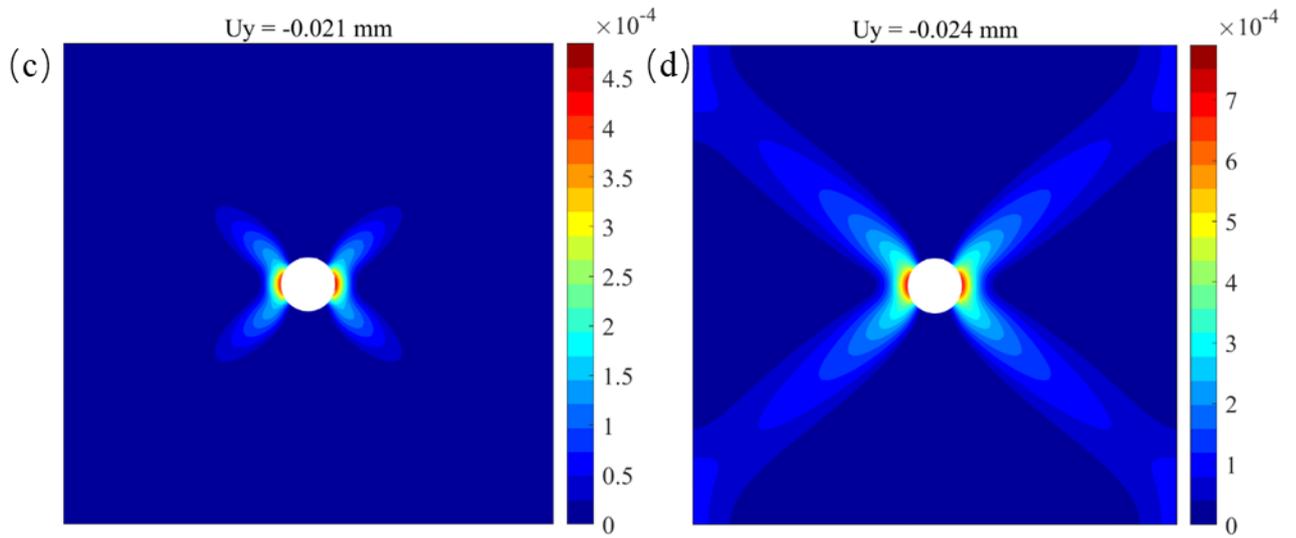

**Fig. 22.** Equivalent plastic strain fields during loading.

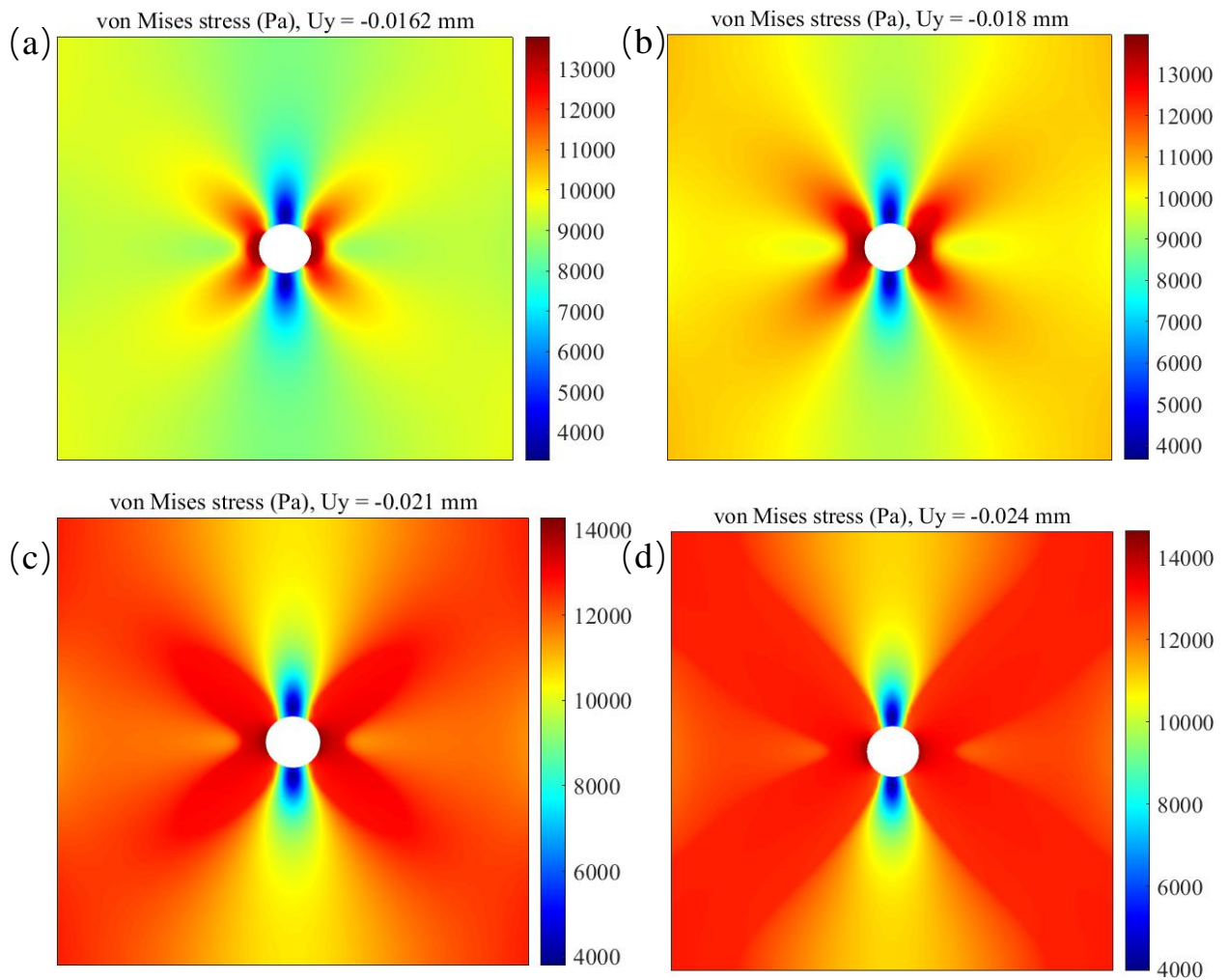

**Fig. 23.** Contour of von Mises stress during loading.

## 6. Conclusion

This study presents an improved NOSB-PD to simulate geomaterials' mechanical response and fracture behaviours. By implementing a divergence formulation of the non-local differential operator and introducing a novel "loading-balance-correction" algorithm, the improved PD model effectively addresses two critical limitations of conventional NOSB-PD:

the surface effect and the fictitious yielding. The integration of a modified hyperbolic-hardening Drucker-Prager model enhances the framework's capability. Validation through comparative analysis with finite element method results and experimental data confirms that the improved NOSB-PD eliminates numerical errors and accurately captures geomaterials' elastoplastic deformation and progressive failure.

**Declaration of Competing Interests**

The authors declare that they have no known competing financial interests or personal relationships that could have appeared to influence the work reported in this paper.
**Acknowledgement**
The authors gratefully acknowledge Professor Haitao YU and Dr. Xiaokun Hu from Tongji University for their conceptual contributions to this work.

**Data availability**

Data generated or analysed during this study are available from the corresponding author upon reasonable request.

**Appendix. Stress integration strategy: closest point projection method (CPPM)**

For those material points entering the elastoplastic deformation stage, the present work uses a backward Euler method for the stress integration, i.e., S-space formulated closest point projection method (CPPM). In the framework of hyperbolic hardening, the Drucker-Prager model presented in the paper, elastic predictor yields:

$$I^{trial} = I^n + 3K\Delta\varepsilon_{ii} \tag{A1}$$

$$\mathbf{s}^{trial} = \mathbf{s}^n + 2G\Delta\mathbf{e} \tag{A2}$$

$$J^{trial} = \sqrt{\frac{1}{2}\left(s_{ij}^{trial} : s_{ij}^{trial}\right)} \tag{A3}$$

where superscript *trial* indicates the trial variables, $K$ is the bulk modulus, $G$ is the shear modulus, $\mathbf{s}$ is the deviatoric stress tensor, $\mathbf{e}$ is the deviatoric stress tensor and $I$ and $J$ are the first stress invariant and the second deviatoric stress invariant, respectively.

The updated $I$ and $J$ can be obtained using the Euler equations:

$$I^{n+1} = I^{trial} - 3K \cdot \Delta\lambda \cdot r_{ii}^{n+1} \tag{A4}$$

$$\mathbf{s}_{n+1} = \mathbf{s}_n^{trial} - 2G \cdot \Delta\lambda \cdot \mathbf{r}_{n+1}^d \tag{A5}$$

where $\mathbf{r} = \frac{\partial g}{\partial \boldsymbol{\sigma}}$, $g$ is the plastic potential function, $r^d$ is the deviatoric part of $\mathbf{r}$ and $\Delta\lambda$ is the plasticity multiplier. The expression of $r$ can be calculated using the chain rule:

$$r_{ij} = \frac{\partial g}{\partial I}\delta_{ij} + \frac{\partial g}{\partial J}\frac{s_{ij}}{2J} = -m\alpha d_0 \delta_{ij} + \frac{s_{ij}}{2J} \tag{A6}$$

leading to:

$$r_{ii}^{n+1} = -m\alpha d_0 \tag{A7}$$

$$\mathbf{r}_{n+1}^d = \frac{\mathbf{s}_{n+1}}{2J_{2,n+1}} \tag{A8}$$

Combining Eq. (A4) to Eq. (A8), the updated $I_{n+1}$ $J_{n+1}$ and $\mathbf{s}_{n+1}$ can be obtained by:

$$I^{n+1} = I^{trial} + 9K \cdot \alpha d_0 m\Delta\lambda \tag{A9}$$

$$J^{n+1} = J^{trial} - \Delta\lambda G \tag{A10}$$

$$\mathbf{s}_{n+1} = \mathbf{s}_n^{trial} - \frac{1}{J_{n+1}}G \cdot \Delta\lambda \cdot \mathbf{s}_{n+1} \tag{A11}$$

Furthermore, Eq. (A11) leads to:

$$\mathbf{s}_{n+1} = \frac{\mathbf{s}_n^{trial}}{1 + \frac{G \cdot \Delta\lambda}{J_{n+1}}} \tag{A12}$$

The plastic internal variable (PIV) and the hardening parameter can be calculated by:

$$\xi_{n+1}^{eq} = \xi_n^{eq} + r_{n+1}^d \Delta\lambda = \xi_n^{eq} + \frac{\Delta\lambda}{2} \tag{A13}$$

$$m_{n+1} = m_0 + \frac{h(1-m_0)\xi_{n+1}^{eq}}{1-m_0 + h \cdot \xi_{n+1}^{eq}} \tag{A14}$$

Once the plastic multiplier Δλ is determined, both the stress tensor and the PIV can be updated. In the S-space formulated CPPM, Δλ is computed by establishing a cubic equation:

$$B_1 \cdot \Delta\lambda^3 + B_2 \cdot \Delta\lambda^2 + B_3 \cdot \Delta\lambda + B_4 = 0 \tag{A15}$$

in which,

$$\begin{aligned}
B_1 &= -\alpha_{14}^2 G - \alpha\alpha_{12}\beta_{11} \\
B_2 &= -2\alpha_{13}\alpha_{14}G + \alpha_{14}^2 J^{trial} - \alpha\alpha_{11}\beta_{11} - \alpha\alpha_{12}\beta_{12} - k\alpha_{12}\alpha_{14} \\
B_3 &= 2\alpha_{13}\alpha_{14}J^{trial} - \alpha_{13}^2 G - \alpha\alpha_{11}\beta_{12} - \alpha\alpha_{12}\beta_{13} - k\alpha_{11}\alpha_{14} - k\alpha_{12}\alpha_{13} \\
B_4 &= \alpha_{13}J^{trial} - \alpha\alpha_{11}\beta_{13} - k\alpha_{11}\alpha_{13}
\end{aligned} \tag{A16}$$

Parameters are:

$$\begin{aligned}
\alpha_{11} &= m_0(1-m_0) + h\xi_n^{eq} \\
\alpha_{12} &= \frac{h}{2} \\
\alpha_{13} &= 1 - m_0 + h\xi_n^{eq} \\
\alpha_{14} &= \alpha_{12} = \frac{h}{2} \\
\beta_{11} &= 9K\alpha d_0 \alpha_{12} \\
\beta_{12} &= 9K\alpha d_0 \alpha_{11} + I^{trial}\alpha_{14} \\
\beta_{13} &= I^{trial}\alpha_{13}
\end{aligned} \tag{A17}$$

The solution for Δλ is a non-negative real number. Finally, the updated stress tensor is given by:

$$\boldsymbol{\sigma}_{n+1} = \boldsymbol{s}_{n+1} + \frac{1}{3}I_{n+1}\boldsymbol{I} \tag{A18}$$

Eq.(A13) and Eq.(A14) compute the equivalent plastic strain.